# Routing Emergency Vehicles in Arterial Road Networks using Real-time Mixed Criticality Systems*

Subash Humagain and Roopak Sinha
*School of Engineering, Computer and Mathematical Science*
*Auckland University of Technology, Auckland, New Zealand*
subash.humagain@aut.ac.nz, roopak.sinha@aut.ac.nz

## Abstract

*Reducing the response time of Emergency Vehicles (EVs) has an undoubted advantage in saving life and property. Implementing pre-emption can aid in achieving it. EVs get unobstructed movement via pre-emption, usually by altering traffic signals and giving a green wave throughout the route. This approach of absolute pre-emption effects adversely on regular traffic by imposing unnecessary waiting. In this paper, we propose a novel emergency vehicle pre-emption (EVP) algorithm implemented in the Vehicular Ad-hoc Network (VANET) that can reduce the imposed undesirable waiting time, but still ascertains EVs meet target response time. We introduce mixed-criticality real-time system scheduling concept where different level of emergencies is mapped with different criticality levels and assign certain success assurance level to respective criticality. We implemented the EVP algorithm for an arterial traffic network and leveraged the use of valuable information that gets transmitted via VANET to make critical decisions. The proposed algorithm can significantly reduce the average waiting time of regular traffic. It also ascertains all EVs with different level of criticality meet target response time respective to their assurance level.*

## 1. INTRODUCTION

Reducing the response time of Emergency Vehicles (EVs) has an enormous impact on saving life and property. According to a study conducted by RapidSOS in the USA, every single minute of delay in response time increases mortality by 1% and incurs 7 billion dollars of extra healthcare expenses [1]. To mitigate this issue, governments impose target response times for Emergency Management Services (EMS). For instance, 90% of critical emergency calls must be responded to within 9 minutes in the USA [2], while in the UK 75% of such cases must be responded to within 8 minutes [3]. For Australia and New Zealand, the target is to respond to 50% of emergency calls within 8 and 10 minutes, respectively [4], [5]. Due to increasing pedestrian population and congested road networks it has become an increasingly difficult challenge for EMS to meet contractual timings.

Researchers and industry practitioners have examined the problem of reducing response time for EVs extensively and have proposed or implemented different solutions. Current solutions offer either used route optimization, signal preemption or both techniques to reduce the response time of EVs [6]. Route optimization selects the fastest route for EVs within the available circumstances. Traffic pre-emption modifies traffic flow to prioritize selected vehicles like EVs. The most widely used pre-emption technique is altering traffic signals to halt normal traffic flow and provide passage to EVs [7]. Present pre-emption techniques use GPS, localized radio, acoustic or line of sight sensors to activate preemption. Emergency Vehicle Pre-emption (EVP) has aided in reducing the response time of EV and served in saving the life from fatalities. Kamalanathsharma and Hancock in their study have shown that EVP can achieve savings of up to 31% in travel times compared to the system without EVP [8]. Though the performance of EVP in reducing EV response time is compelling, its consequences over general traffic (unnecessary waiting time) cannot be overlooked.

A Vehicular Ad-hoc Network (VANET) can aid in optimizing the time of triggering and ceasing of EVP. VANET is the wireless ad-hoc network created by moving vehicles where vehicle communicate with other vehicles or infrastructure using dedicated short-range communication standards outlined by IEEE 802.11p [9]. In VANET roadside infrastructures like traffic controllers and vehicles can share valuable information like speed, position, lane, route, and time of arrival, which are fundamental in optimizing green time available for EVs. This, in turn, can reduce the negative effects EVP may have over general traffic. Several studies have implemented VANET in sharing the current position of EV to determine the right time to trigger and cease EVP [10], [11], [8]. Unibaso et al. used standard Cooperative Awareness Message (CAM) defined by the European intelligent transportation system in their traffic control algorithm to provide a green light to EVs[12]. Similarly, Walz and Behrisch also used CAM to provide a green wave to EV travelling through multiple intersections [13]. Jordan and Cetin implemented VANET for efficient passage of EVs in closely spaced traffic intersections by preempting traffic signals in a definite order so that existing traffic within these intersections can be discharged prior to the arrival of EV [14]. In [15], a dynamic traffic scheduling algorithm was designed that can fine-tune the green time allocated for emergency vehicles so that there is a lesser impact on normal traffic.

pre-emption techniques have been studied widely for EV routing, but EMS companies are always struggling to achieve the target response times. This is because almost all cutting edge research and implementations of pre-emption techniques focus on reducing the travel time of an individual EV in an optimized route providing that EV with the highest priority. But in real-life city traffic and in case of natural calamities, there can be multiple EVs servicing different levels of emergencies at the same time. The approach of prioritizing a single EV can reduce the travelling time of that particular EV, but does not assist the EMS in meeting the target response times for all EVs on the ground. Moroi and Takami have also pointed towards the limitations of prioritizing single EVs in case of emergencies where we may require to route multiple EVs at the same time [16]. In such cases, other EVs following a recently prioritized EV experience more congestion. Furthermore, providing absolute priority to an individual EV serving any level of emergency increases the overall waiting times for normal traffic. This problem can be solved if multiple EVs serving within a particular time are assigned with different levels of priority and EVP is performed accordingly.

In this study, we propose an EVP technique for multiple EVs with different levels of priority. We leverage the use of VANET for transmitting critical information like current position, speed, the time EV takes to pass the intersection, and the route the EV follows in deciding when to trigger the EVP. Important decision parameters like speed, position and route of vehicles are easier to access using VANET which is impossible from traditionally used inductive loops [17]. There can be conflict in prioritizing EVs when they share the same route or some common intersections within that route. This issue is solved by assigning different levels of priorities to EVs. Assigning a different level of priorities to EVs is a common practice in EMS industry, for example, St. John's in New Zealand classify serious life-threatening incidents as purple, red and, and less serious emergencies as orange. Without affecting the generality of our work, we use the New Zealand system for illustration of concepts presented in this paper.

The EVP technique proposed is novel as it provides a way to arbitrate multiple EVs in an arterial road network. An arterial road network is the backbone of any urban road that is usually used to transport traffic from small collector roads to expressways or motorways. Rather than implementing EVP in a single intersection, modelling and implementing EVP for an arterial road network allows for real-world visualization and more efficient EV routing. Performance parameters like waiting times and throughput achieved from such implementations are realistic and will be helpful for EMS to evaluate their actual performance. In addition, it also helps traffic engineers and transport planners to anticipate the widespread effect of EVP over general traffic flows rather than limiting such explorations to a single intersection. Moreover, adding multiple EVs into the EVP model can solve unsolved problems faced by EMS in prioritizing EVs when two or more of them have conflicts passing an intersection.

The proposed traffic control algorithm designed for EVP of multiple vehicles is constructed using a novel approach of mapping traffic control domain into Real-Time MixedCriticality Systems (RTMCS) task scheduling. In real-time systems, the accuracy of jobs being processed does not rely solely on correct functionality but also on the timely completion of tasks and computations. In RTMCS, jobs with multiple levels of criticality like mission-critical, non-critical and safety-critical are designated with different timing constraints. The system is considered a failure if it cannot meet these timing values [18]. We map EV priority levels (purple, red and orange) with different levels of criticality in RTMCS and impose EVP system to ascertain a certain level of success assurance against a different level of criticality. The arbitration of conflicts that arise from scheduling equal priority EVs through the same intersection is equivalent to the well-known problem of scheduling with conflicts in real-time systems and can be resolved using conflicts graphs. Whenever two or more EVs must pass via a single traffic intersection, they cannot be scheduled simultaneously if they are travelling in conflicting directions. This problem of resource sharing in real-time systems is dealt through modelling a conflict graph [19]. We model and utilize conflict graphs to create an optimal schedule for EVs in traffic intersections. As EV arrival at an intersection is a random event, this paper proposes an *online* scheduling algorithm. The implementation of the proposed EVP in an arterial traffic network for multiple EVs with different levels of priorities shows promising results. We conducted multiple experiments using the microscopic traffic simulator Simulation of Urban Mobility (SUMO) [20]. Simulation results show that the waiting times and overall travel times are significantly reduced when EVP is enabled, as compared to traditional static traffic light control. Moreover, the EVP algorithm seems to also reduce waiting times for non-EV traffic when compared to absolute pre-emption.

## 2. SYSTEM MODEL

A four-legged isolated traffic intersection with adaptive traffic control system implemented in a VANET environment can be embedded into an arterial road network with multiple intersections. Let us consider a four-legged traffic intersection designed for left-hand driving as illustrated in the left side of Fig. 1. It consists of four road segments from North, South, East, and West directions. Each road segment is divided into two approaches so that this intersection has a total of eight approaches. Every single approach has conflict in movement with five other approaches and has only one non-conflicting approach. For example, traffic from approach 2 cannot be scheduled with traffic form approaches 3, 4, 5, 7 and 8, but it can still be scheduled with traffic from approaches 1 and 6.

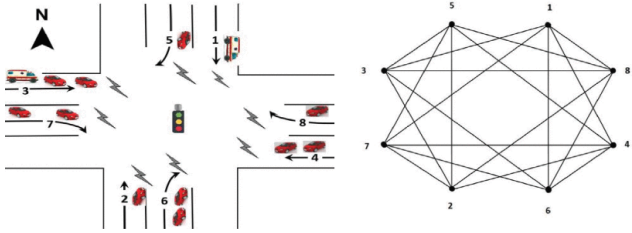

Fig. 1. Four-way traffic intersection and its equivalent conflict graph

A conflict graph is used to model concurrency in simultaneous operations in real-time systems and can be used in traffic intersections. A traffic signal must always schedule non-conflicting approaches together. This can be analogically mapped to task scheduling in a real-time system which can be implemented through conflict graphs. A conflict graph with nodes N and edge E is represented as $G = (N, E)$ where each approach from traffic intersection is a node (in $N$) and an edge (in $E \subseteq N \times N$) is a connection between two conflicting approaches as depicted in the right side of Fig. 1. This approach can be generalized to any number or types of approaches. Our traffic scheduling algorithm schedules traffic by picking a set of non-conflicting approaches at every instance and allowing it appropriate green time.

VANET provides precious data to activate efficient EVP. VANET is a subset of Mobile Ad-hoc Network (MANET) designed especially for vehicular communication. The network topology of MANET changes rapidly as nodes are permitted to move in any direction. Routing protocols for MANET are unable to provide optimum throughput for complex trails created by rapid change in relative positions of moving nodes due to the high speeds of vehicles in a predefined road. Therefore, VANET has become the technology of choice in such circumstances [21]. In this study, we assume that all EVs are equipped with Dedicated ShortRange Communication (DSRC) devices for the vehicle to infrastructure communication and GPS to collect speed and position information. Information about the route EVs follow from origin to the destination is initially set by the routing application installed within the vehicle. The communication between EVs and the road-side traffic controller is carried out in 5.9 GHz spectrum bandwidth for DSRC as standardized by IEEE 1609.4 protocol. The traffic controller receives beacons transmitted from EVs, which contain information like speed, position and route of an EV. Knowing the route before the activation of EVP allows the controller to decide which approach must be allowed to provide EVs right of way. Speed and position data allows determining exactly when to switch the current traffic signal phase to green or by what time the current green phase should be extended. The current speed of EV replicates dynamic traffic behaviour of that particular lane, and for example, a congested lane needs the controller to reserve a longer green time than a free lane.

A single intersection EVP approach is insufficient. Calculation of exact green time required for an EV to pass through an intersection is very crucial as allowing unnecessarily extended green time negatively affects the flow of general traffic waiting to pass through the intersection. In order to ensure that such issues are minimized, we calculate the exact time the traffic controller needs to reserve for the EV in real time so that it can pass the intersection within this time using information like current speed and position transmitted by the EV. Allowing a well-timed green signal for a particular intersection still cannot guarantee that the EV can meet the contractual time to reach a destination. So we have designed our system to work on an arterial road network where an EV has a source and a destination. The fastest route is initially assigned to an EV. This route contains multiple intersections and a distance to cover. The contractual target time can only be ascertained if the EV maintains the maximum allowed speed throughout the journey. The main aim of the EV traffic control algorithm installed in each traffic controller is to allow the EV to maintain this speed, as much as possible. We take advantage of the information beacon transmitted by EVs to maximize this possibility. Additionally, to replicate the real-world situations, we introduce multiple EVs in an arterial network at the same time and study their routing.

Assigning a different level of criticality to EVs and implementing EVP accordingly is less disruptive to normal traffic. When EVs are serving emergencies, they travel with sirens and lights on. Current EVP systems provide a green signal to EVs when they identify an EV's presence through GPS, localized radio, acoustic or line of sight sensors. Once the presence of EVs is confirmed, they get absolute priority throughout the route. This is more disruptive to normal traffic, especially at intersections. In most cases, all EVs are not serving to the same levels of emergency. So instead of providing absolute priority at all intersections we can use a mixed model of EVP. In our experiment, we have assigned three levels of criticality to EVs. Our algorithm now schedules EVs with different levels of criticality and ascertains the times they reach the destination and if this time can be within a set response time. In doing so, the algorithm identifies the intersections it needs to preempt and leaves other intersection unaffected. This approach massively reduces the effect of EVP on normal traffic.

Scheduling EVs with different levels of criticality is analogous to RTMCS. In RTMCS the calculation of WorstCase Execution Time (WCET) depends on the criticality of the task. For example, safety-critical tasks have lowest WCET than mission-critical tasks, and a non-critical task has the highest WCET [22]. This aligns exactly with our case of multiple EVs with different levels of criticality. Standard response times of 8, 12 and 16 minutes for high, medium and low critical cases of emergencies imply medium and low criticality cases are respectively assigned response times that are 1.5 and 2 times higher than high criticality cases. We consider these factors as assurance levels $L_i = \{l_1, l_2, l_3, ..., l_n\}$ be a set of assurance levels, which is defined for each EV . In this study, we define target travel time at any instant $i$ as $Tt_i$ for an EV as follows.

$$Tt_i = (D_i/V_{max}) * L_i \qquad (1)$$

Where $D_i$ is the distance between the current EV location and the target location and $V_{max}$ is the speed limit of the particular lane. We also define current travel time for any instant $i$ as $Tc_i$ and calculated as

$$Tc_i = (D_i/V_{EV}) \quad (2)$$

Where $V_{EV}$ is the current speed of EV. Also, the relative difference in time $\Delta T$ is calculated as

$$\Delta T = Tc_i - Tt_i \quad (3)$$

From equation. 3 if the $\Delta T \leq 0$ then the EV can reach its destination within its target travel time without triggering EVP. When $\Delta T > 0$, it implies the route of EV is congested and EV cannot reach the destination within its target travel time and hence we may need to trigger EVP immediately.

The overall working of the system is depicted in Fig. 2. Consider the situation where we have three EVs, EV1, EV2 and EV3, serving at the same time with three different levels of criticality. This algorithm can perform equally well for more than three EVs and more than three levels of criticality.

Initially, all the EVs are assigned with routes. Whenever an EV approaches an intersection, it sends information about its route (origin and destination), current position, criticality level and speed to the traffic controller. The traffic controller first checks if there is any other EV approaching the same intersection. If there is more than one EV approaching the same intersection controller checks the level of criticality of EV and processes the request of EV with higher priority else it processes the current request. It then calculates the value of $\Delta T$ and determines the requirement of EVP. When pre-emption is required, the controller calculates the reserve green time required for each EV to pass the intersection from the position data. If the current state of the traffic light is green, it extends current green time with reserve green time else it alters the current phase and provides green phase to EV. The EVP algorithm is shown in Algorithm 1. Parameters like $EV_i.position$, $EV_i.speed$, $EV_i.criticality$, and $EV_i.destination$ can be set as vehicle parameters SUMO and are accessible during the simulation.

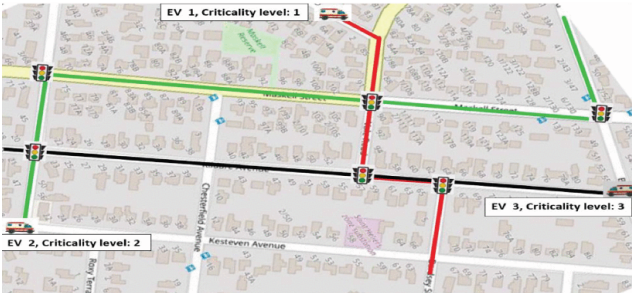

Fig. 2. System Model with multiple intersections and EVs

---

**Algorithm 1** Emergency Vehicle pre-emption
**Input:** Emergency vehicles $EV_i(1,2,....,n)$
**Output:** Alter traffic signal
1: **for** each $EV_i$ **do**
2:     find $EV_{min}$ with minimum $EV_{min}.criticality$;
3:     $\Delta T$ = calculate $\Delta T(EV_{min})$ ;
4:     **if** ( $\Delta T \leq 0$) **then**
5:        Reserve time = calculateReservetime($EV_{min}$);
6:        action = alterTrafficLight(Reserve time);
7:     **else**
8:        No action;

## 3. IMPLEMENTATION

This section illustrates the implementation of the proposed EVP algorithm using SUMO, OMNET++ and VEINS. SUMO is a widely-used and open-source realistic microscopic traffic simulator. OMNET++ is a simulation library designed to implement different network simulations. We use VEINS as a platform to couple SUMO and OMNET++ to simulate vehicular communication using IEEE 802.11p. FullDuplex communication is established between SUMO and OMNET++ by VEINS using Transmission Control Protocol (TCP), which helps in examining how VANET implementations affect road networks [23].

Simulation parameters in SUMO can be altered using TRACI when the simulation is live. SUMO gathers prior information from its different components and runs the simulation continuously until it produces the result. The parameters of the simulation cannot be altered during the course of the simulation. To access the simulation parameters and alter its course, we use an extension tool developed in Python called TRACI. Different applications can communicate and control SUMO during simulation via client-server architecture using TRACI. All the communicating applications are set as client and SUMO acts as a server. They use TCP sockets to communicate with each other using TCP/IP protocol. Client applications send TCP packets to SUMO requesting to alter or change parameters like altering traffic light logic, changing vehicle route, lane, destination etc. In response to the client's request, SUMO alters these parameters and replies back to the client application. TRACI also helps to establish communication between traffic simulator SUMO and the network simulator OMNET++.

SUMO and OMNET++ run concurrently. Each vehicle in SUMO is treated as a moving node in OMNET++. A mobile node gets created once a vehicle appears in the road network and disappears when the vehicle reaches its destination. Every new node created in OMNET++ has a unique MAC ID as a vehicle in SUMO has $vehicle\_ID$. We can use information transmitted over the network and make preemptive decisions and implement it by changing the traffic light signal phases.

### A. Simulation parameters
Realistic traffic parameters are utilized for simulation based validation of the proposed solution. We conduct our experiment over a section of Auckland city's arterial road network and calibrate the network with vehicular data. All intersections are initially modelled with static Traffic Light Control (TLC) logic. Traffic saturation is set to 1800 pcu/hr for every lane. Vehicles arrive randomly following the Poisson distribution. To achieve variable vehicular flows in a particular lane, we assign probability values for arrival rates. The car-following model is set to Krauss and parameter sigma is set to 0.5. For EVs vehicle class ($vClass$) is set to "emergency". We use New Zealand-based emergency codes purple, red and orange for assigning high, medium and low levels of criticality to EVs, respectively. Once EVP is activated TRACI controls the simulation and alters traffic light phases to provide green waves for EVs as required.

The primary aim of our algorithm is to ascertain that EVs with different levels of criticality reach their destinations within desired times. EVs are created randomly with randomly assigned criticality. We change the traffic volumes in the route of EV from low (450 pcu/hr/ln), medium (850 pcu/hr/ln) and high (1800 pcu/hr/ln) and for each setting, we perform simulations multiple times. In all cases, the EVs are generated randomly, assigned with random routes and criticality. According to an annual report published by St. John New Zealand, out of 546,000 emergency calls 63.6% cases were highly critical, 23.2% cases were medium and the remaining 13.3% of the cases were low critical [4]. In our experiments as well, we have maintained the same proportions of the levels of criticality. We have considered 1% of total vehicles generated as EVs for a simulation time of 10,000 seconds and with longer duration of simulation time a lesser percentage of EVs can be considered. We measure the number of times EVs met the target travel time of 8, 12, and 20 minutes for purple, red and orange level of criticality respectively in terms of the success rate of EVP. We also measure the average waiting time of both non-EVs and EVs, queue length, and overall throughput in a system implemented with our algorithm and compare it with systems without pre-emption and absolute pre-emption (EV receives green phase once detected near intersection until it leaves).

## 4. PERFORMANCE EVALUATION

Our EVP implementation shows promising results when compared to traditional systems containing either no preemption or absolute pre-emption. As claimed in earlier sections, the impact of pre-emption is huge in normal vehicles. The claim made by other pre-emption technique that has been implemented in a single intersection cannot justify a real-world situation where multiple EVs move from source to the destination covering multiple intersections. The cost incurred by increased waiting times of normal vehicles when pre-emption is calculated in terms of $CO_2$ emissions and the corresponding price of fuel used, which are high for the traditional settings and relatively lower in the proposed EVP implementation. Our experiments study five important metrics: the percentage of EVs meeting their target travel time successfully, average waiting times for both EVs and non-EVs, average queue lengths at intersections and overall throughput of the network.

Fig. 3 shows the percentage of success EVs achieve with the proposed EVP algorithm in low, medium and high traffic densities. Whenever an EV is lagging behind to meet its target travel time, pre-emption is activated, which means that success percentage is expected to be 100%. However, after taking into account traffic uncertainties as modelled in SUMO, the overall success percentage is lower, especially in peak traffic. All simulations reported here ran for 36000 seconds, and in some cases, a longer duration of simulation gives almost perfect results. Our result depicts that at higher traffic densities few EVs struggle to meet target response time but still maintain promising 96% success. Some EVs with a high level of criticality, miss target response time, as their number is very high as compared to EVs with other criticality levels but still maintain exuberant 95% success.

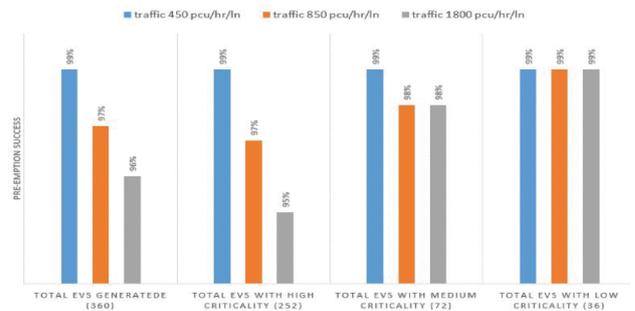
Fig. 3. Success percentage of EVs meeting target travel time

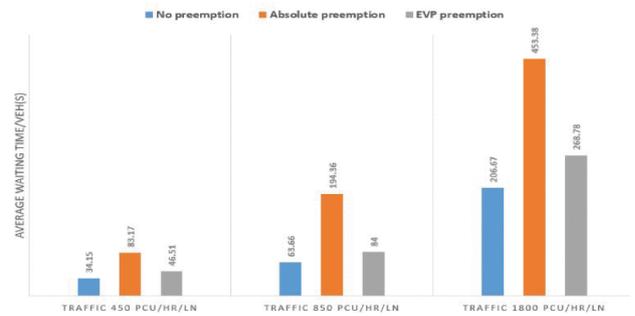
Fig. 4. Average waiting time of non-EVs

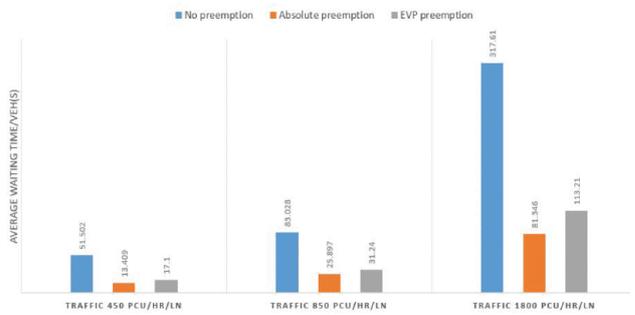
Fig. 5. Average waiting time of EVs

Pre-emption affects normal traffic by increasing their average waiting times. Absolute pre-emption incurs a very high waiting time for non-EVs resulting in huge financial losses. A system without pre-emption becomes an obstacle to EVs. We implemented a practical approach where EVs are assigned with different level of criticality and imposed different target travel times. The EVP algorithm ascertains that EVs meet these target travel times. At different traffic penetration rates, the average waiting time of non-Evs is reduced drastically, almost comparable to the system without pre-emption. This ensures that the implementation of EVP algorithm can have huge perennial savings in terms of fuel cost and $CO_2$ emission. The results are depicted in Fig. 4. Since we used a mixed model in making decisions to activate pre-emption, EVs routed using EVP algorithms experienced slightly increased waiting time as compared to a system with absolute pre-emption but performed exceptionally well than the system with no pre-emption as shown in Fig 5.

Queue length for an intersection measures the efficiency of any traffic control algorithm. It is measured as the average number of vehicles waiting at an intersection. The EVP algorithm's impact on this metric is shown in Fig. 6. The experimental results show that average queue length is reduced up to 36% using EVP algorithm as compared to absolute pre-emption and still is comparable to the system with no pre-emption.

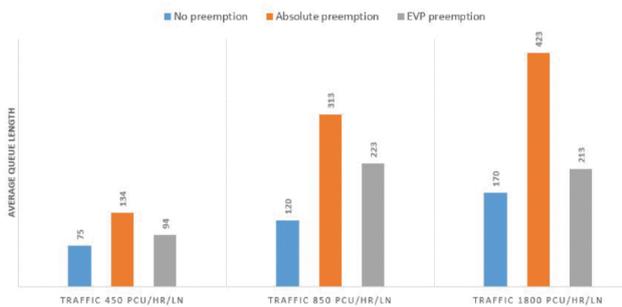

Fig. 6. Queue length comparisons

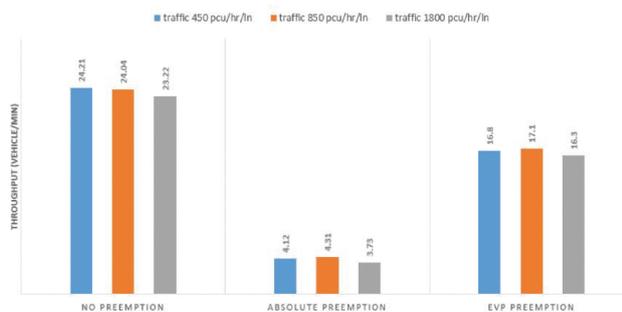

Fig. 7. Throughput comparison

The total traffic volume an intersection can flush through is measured in terms of throughput. This well-known productivity indicator is used to compare EVP, absolute preemption and no pre-emption in Fig. 7. The results illustrate that the throughput of EVP implemented system is high and is comparable to the system without pre-emption. The system with absolute pre-emption has the lowest throughput.

## 5. CONCLUSION AND FUTURE WORK

In this study, we try to solve ever-existing two problems caused by the pre-emption of emergency vehicles (EVs). First, though pre-emption techniques have been studied extensively, emergency management service providers always struggle to meet target response time. Second, a substantial cost of pre-emption that normal traffic has to handle in terms of waiting. We postulated a novel emergency vehicle pre-emption (EVP) algorithm implemented in Vehicular adhoc network that aid to solve these issues. We introduced different levels of criticality for different levels of emergencies and assigned a certain level of success assurance in terms of target travel time for these criticality. Unlike other studies, instead of implementing EVP algorithm in a single intersection, we implemented it in an arterial traffic network. We ran exhaustive simulations, and the results indicate that EVP algorithm can significantly reduce the average waiting time of normal traffic but still assures all EVs meet their target response time. In future, we aim to implement EVP algorithm in a larger city map with realworld calibrated traffic data-set. Also, we intend to see the VANET parameters like message delay range, beacon rate and throughput optimization.